\documentstyle[12pt]{article}
\begin{document}
\baselineskip 7mm
\title{ How to Measure the Fractal Geometry of the Relativistic Fermion
Propagator }
\author{H. Kr{\"o}ger \\ [2mm]
{\small\sl D{\'e}partement de Physique, Universit{\'e} Laval,
Qu{\'e}bec, Qu{\'e}bec, G1K 7P4, Canada } \\
{\small\sl Email: hkroger@phy.ulaval.ca } }
\date{April 1995 \\
Universit\'e Laval preprint: LAVAL-PHY-3/95}
\maketitle
\begin{flushleft}
{\bf Abstract}
\end{flushleft}
We study the geometry of propagation of relativistic fermions. We propose how
to measure its quantum mechanical length. Numerical lattice results for the
free propagator of Dirac-Wilson fermions yield Hausdorff dimension $d_{H}=2$
for the unit-matrix component and $d_{H}=1$ for any $\gamma$-matrix component.
A possible generalization when matter interacts with radiation is discussed. \\
PACS index: 03.65.-w, 05.30.-d
\setcounter{page}{0}

\newpage
\begin{flushleft}
{\bf 1. Introduction} \\
\end{flushleft}
What do we know about the geometry of propagation of massive particles in
non-relativistic quantum mechanics? According to Feynman and Hibbs
\cite{kn:Feyn65} quantum mechanical paths are zig-zag lines, which are no-where
differentiable, exhibiting self-similarity when viewed at different length
scales.
In terms of modern language this is a fractal curve \cite{kn:Mand83}.
Abbot and Wise \cite{kn:Abbo81} have shown for free motion that quantum
mechanical paths are fractal curves of Hausdorff dimension $d_{H}=2$. Actually,
in quantum mechanics the concept of paths is not well defined.
If, however, one goes over from real to imaginary time the path integral
becomes
mathematically well defined (Wiener measure). In imaginary time (Euclidean)
quantum mechanics the free motion resembles the Brownian motion of a classical
particle \cite{kn:Nels66}.
Then one has a stochastic interpretation as classical Brownian motion and paths
can be considered as random variables of a Gaussian process \cite{kn:Roep94}.
The typical path of a classical particle
carrying out a Brownian motion is a fractal curve with $d_{H}=2$. According to
Itzykson and Drouffe \cite{kn:Itzy89} when following a Brownian curve on a
lattice from $\vec{x}_{0}, t_{0}$ to
$\vec{x}_{1}, t_{1}$ the typical distance behaves as
\begin{equation}
| \vec{x}_{1} - \vec{x}_{0} | \sim (t_{1} - t_{0})^{\nu}, \;\;\; \nu=1/2,
\label{eq:Brownian}
\end{equation}
and $\nu$ is called a critical exponent. (by analogy with the power laws in the
theory of critical phenomena).
Considering the typical distance as a function of the elementary length $\Delta
x$ then defines the Hausdorff dimension  (see Eq.(2) below). It is important to
note that $d_{H}=2$ for free motion in quantum mechanics does not change when
going over from real time to imaginary time.

\bigskip

For the case of a massive interacting quantum mechanical particle
Campesino-Romeo et al. \cite{kn:Camp82} have shown analytically $d_{H}=2$ for
paths of the harmonic oscillator.
Other potentials have been investigated by numerical simulations on the lattice
\cite{kn:Krog95}.
It turned out that $d_{H}=2$ for local potentials (harmonic oscillator, Coulomb
potential).
However, $d_{H} \neq 2$ was found for velocity-dependent interactions. Such
velocity-dependent actions are supposed to play a r\^ole in condensed matter
(dispersion relation of a massive particle
travelling in a medium) or in Brueckner's theory of nuclear matter
\cite{kn:Brue55}.

\bigskip

The Hausdorff dimension of field configurations has recently become a subject
of interest in quantum gravity (in particular in 2-D), where the r\^ole of
fields is played by the geometry of space-time
\cite{kn:Kawa90} - \cite{kn:Catt95}.
The Hausdorff dimension can, e.g., be defined as a power law relation between
two dimensionful
observables at a critical point \cite{kn:Kawa90}. E.g., for quantum gravity in
D=2, one has $d_{H}=4$ \cite{kn:Ambj95}.It is interesting to establish
relations between the Hausdorff dimension and
critical exponents. Such a relation is known to exist between $d_{H}$ and the
critical exponent $\nu$,
which is the analogue of a critical exponent of the spin-spin correlation
function in statistical mechanics. In quantum gravity \cite{kn:Ambj95} $\nu$ is
defined as the critical exponent of a mass,
$m(\Delta \mu) \sim (\Delta \mu)^{\nu}$, when the cosmological constant tends
to its critical value $\Delta \mu \rightarrow 0$, where the mass $m(\Delta
\mu)$ characterizes the fall-off behavior of a two-point function $G_{\mu}(r)
\sim \exp[-m(\Delta \mu)r]$. Then holds the following relation (scaling law)
$d_{H}=1/\nu$, relating the Hausdorff dimension to the critical exponent $\nu$
\cite{kn:Ambj95}.
Thus also $d_{H}$ plays the r\^ole of a critical exponent.

\bigskip

Critical exponents of a theory determine its universality class and thus
classify the theory. The above example of quantum gravity leads us to the
question: Can the Hausdorff dimension play a similar r\^ole for models like,
e.g., theory of matter interacting with radiation, in condensed matter, in the
theory of medium energy nuclear physics (Dirac phenomenology), or in high
energy for the theory of quarks and gluons (plasma)?
A related point of view is the following: In quantum mechanics the Hausdorff
dimension of typical paths may differ from the standard value $d_{H}=2$ in the
presence of velocity-dependent potentials
\cite{kn:Krog95}, also if one considers quantum mechanics in a background
medium corresponding to curved space-time. Such a situation occurs, e.g., when
considering
the relativistic propagation of a particle in nuclear matter (e.g., a neutron
star),
or when considering a relativistic particle impinging on the surface and
propagating a short distance in ordinary matter.

\bigskip

With those questions in mind we want to start out here by asking the much
simpler question: What happens to the geometry of propagation of a massive
particle in relativistic quantum mechanics?
We want to study the question by considering the most simple massive
relativistic particle occuring in nature: the non-interacting Dirac fermion.
The purpose of our work is firstly
to compute $d_{H}$ in general and secondly to suggest a definition based on the
propagator which allows for a generalization to interacting fermions.

\bigskip

\begin{flushleft}
{\bf 2. How to measure the geometry of paths in non-relativistic quantum
mechanics?} \\
\end{flushleft}
The Hausdorff dimension of a fractal curve is defined as follows: Suppose one
has an elementary length scale (resolution) $\Delta x$ to cover the curve.
Experimentally, $\Delta x$ corresponds to the resolution of an experimental
apparatus (yardstick, wavelength of light in microscope).
Measuring the length $L$ of a curve in terms of an elementary length $\Delta
x$, the property of being fractal is captured in the Hausdorff dimension
$d_{H}$, defined by
\begin{equation}
L_{0} \sim_{\epsilon \rightarrow 0} L \epsilon^{d_{H}-1},
\epsilon =\Delta x/L_{1} \label{eq:Hausdorffdim}
\end{equation}
where $L_{1}$ is a fixed length. The Hausdorff dimension $d_{H}$ is a number
chosen such that $L_{0}$ becomes independent of $\epsilon$ in the limit
$\epsilon \rightarrow 0$.
Hausdorff has given a precise definition of "resolution" by covering the cutrve
with $L/\Delta x$ spheres of diameter $\Delta x$.

\bigskip

The definition given by Eq.(~\ref{eq:Hausdorffdim}) has been applied in
Ref.\cite{kn:Abbo81,kn:Krog95} to characterize the typical path of a quantum
mechanical trajectory and in this sense the Hausdorff dimension has been
computed. But in order to be precise one has to say what is meant by length $L$
and resolution $\Delta x$ for quantum trajectories.
Suppose we consider the amplitude for propagation of a particle, being at
$t=0$ in a state characterized by position $x_{in}$ and at $t=T$ in a state
characterized by position $x_{fi}$. We discretize time
$t_{0}=0 < t_{1}< \cdots < t_{N}=T$, with $\Delta t=\delta$ and $T=N\Delta t$.
Then the amplitude is given by a path integral, which after discretization
of time reads
\begin{equation}
\left. Z(\delta) = \int d x_{1} \cdots d x_{N} \;
\exp[i S[x_{k},\delta]/\hbar]\right|_{x_{0}=x_{in},x_{N}=x_{fi}}.
\label{eq:partition}
\end{equation}
We have denoted $x_{k}=x(t_{k})$. This amplitude is an approximation of the
continuum (exact) amplitude, obtained by taking the limit $\Delta t \rightarrow
0$, i.e., $N \rightarrow \infty$ in an appropriate way.
A suitable observable to study the geometry is the propagator length defined by
\cite{kn:Krog95}
\begin{eqnarray}
< L(\delta) > &=&
\left. < \sum_{k=0}^{N-1} | x_{k+1} -x_{k} | >
\right|_{x_{0}=x_{in},x_{N}=x_{fi}}
\nonumber \\
&=& \left. \frac{1}{Z} \int d x_{1} \cdots d x_{N} \;
\sum_{k=0}^{N-1} |x_{k+1} -x_{k}| \;
\exp[i S[x_{k},\delta]/\hbar] \right|_{x_{0}=x_{in},x_{N}=x_{fi}}.
\label{eq:quantumlength}
\end{eqnarray}
In quantum mechanics, the resolution of length is given by the dynamics. It is
natural to define it as average length increment corresponding to a time
increment $\delta$ \cite{kn:Abbo81},
\begin{equation}
< \Delta x  > = \left.\frac{1}{N}
 < \sum_{k=0}^{N-1} | x_{k+1} -x_{k} | > \right|_{x_{0}=x_{in},x_{N}=x_{fi}}.
\label{eq:quantumincrement}
\end{equation}
As fixed length $L_{1}$ one can choose, e.g., the length of the classical
trajectory from $x_{in}, t=0$ to $x_{fi}, t=T$ given by the classical continuum
action.
If one is interested only in $d_{H}$, i.e., the exponent of the power law
Eq.(~\ref{eq:Hausdorffdim}),
then any fixed length $L_{1}$ can be chosen, in particular the length of the
straight line between $x_{in}$ and $x_{fi}$.
In the continuum limit of quantum mechanics one takes the limit $\Delta t =
\delta \rightarrow 0$, which implies $< \Delta x > \sim \epsilon \rightarrow
0$.
The length $< L >$ has been computed by numerical simulations on the lattice
using imaginary time (Euclidean) quantum mechanics and $d_{H}$ has
been extracted \cite{kn:Krog95}.

\bigskip

\begin{flushleft}
{\bf 3. Definition of propagation length in relativistic quantum mechanics} \\
\end{flushleft}
When trying to generalize the above considerations to relativistic quantum
mechanics, one is faced with the following problem: (a) In quantum mechanics,
position of a particle is an observable and can be measured. In quantum field
theory position is not an observable.
(b) The particle number is conserved in quantum mechanics. In quantum field
theory particles can be created and annihilated. (c) In quantum mechanics
particles propagate only forward in time. In quantum field theory particles
propagate forward and backward (anti-particles) in time.
Thus in a relativistic theory one must look at space and time dependence,
corresponding to a causal propagation of a massive particle. From the
mathematical point of view as we have pointed out above, the path integral in
imaginary time quantum mechanics is well defined in terms of a stochastic
process.
This can be generalized to Euclidean path integrals of bosonic (polynomial)
field theory which are mathematically well defined,
allowing an interpretation as stochastic process \cite{kn:Glim81,kn:Roep94}.
The measure gives the dominant contributions of no-where differentiable curves
\cite{kn:Glim81}. However, for fermion field theory, there is not yet a strict
mathematical formulation in terms of a stochastic process,
although Osterwalder and Schrader \cite{kn:Oste72} have established a
Feynman-Kac formula for fermion fields.

\bigskip

{}From those remarks it is evident that the definition of propagator length
given by Eq.(4) for non-relativistic quantum mechanics can not be simply taken
over to relativistic quantum mechanics. The definition (4) is based on a
discretization of time (taking $\Delta t=\delta$ finite and letting $\delta
\rightarrow 0$ in the end when extracting the critical exponent).
In relativistic quantum mechanics we discretize time and space, i.e., we work
on a lattice with some finite lattice spacing $a$ (letting $a \rightarrow 0$ in
the end).
Let us introduce a new definition for the length of propagation for
relativistic quantum mechanics, by considering, in particular, the Dirac
fermion action.
It is well known that the corresponding lattice action is plagued by the
so-called fermion species doubling problem: 16 copies of fermions with the same
mass occur (as poles of the fermion propagator).
A way out commonly used is the Wilson action, which lifts the species doubling
by adding another term. The (Euclidean) Wilson-Dirac action on the lattice
reads
\cite{kn:Mont94}
\begin{equation}
S[\psi,\bar{\psi}] = \sum_{m,n} \bar{\psi}_{m} K_{m,n} \psi_{n}.
\end{equation}
Here $m, n$ are indices (tupel) which denote lattice sites (e.g.,
$x_{n} = n a$, $n = 0, \pm 1, \dots, $ in $D=1$).
The fields have been rescaled
$a^{3/2}(a m + 4 r )^{1/2} \psi \rightarrow \psi$,
$\kappa = 1/(2 m a + 8 r)$
is the hopping parameter and $r$ is the Wilson parameter (usually chosen to be
$r=1$).
The matrix $K$ is expressed in terms of the hopping matrix $M$
\begin{eqnarray}
K_{m,n} &=& \delta_{m,n} - \kappa M_{m,n},
\nonumber \\
M_{m,n} &=& \sum_{\mu =1}^{4} (r + \gamma_{\mu}) \delta _{m+\hat{\mu},n}
+ (r-\gamma_{\mu}) \delta_{m -\hat{\mu},n}.
\end{eqnarray}
Here $m+\hat{\mu}$ denotes the lattice site next to site $m$ in the positive
$\mu$-direction.
The fermion propagator is given by
\begin{equation}
< \psi_{n} \bar{\psi}_{m} > = (K^{-1})_{n,m}.
\end{equation}
The matrix $M$, coming from the kinetic term of the fermion action and from the
Wilson term, allows the fermion to hop from one lattice site to the next
neighbour lattice site.
We suggest to define the length of the fermion propagator by counting in a
non-perturbative way the number of hoppings.
In particular we suggest the following definition
\begin{equation}
< L_{m,n} > = \frac{ \partial \log <\psi_{n} \bar{\psi}_{m} > }
{ \partial \log \kappa }.
\end{equation}
The indices $m ,n$ denote the lattice sites, where the fermion is created and
annihilated, respectively. By expanding the right-hand side of Eq.(9) as a
power
series in the hopping parameter, one finds that the $p$-th power of the hopping
matrix $M$, which allows the fermion to hop a distance $pa$, gets multiplied
with a factor $p$. Thus $< L_{m,n} >$ can be interpreted as a
(non-perturbative) counter of how many times a fermion hops between sites $m$
and $n$.

\bigskip

The classical length is $L_{class} = | x_{m} - x_{n} |$. One has to compute
numerically on the lattice $< L >$ as a function
of $a/L_{class}$. The goal is to extract a critical exponent $\gamma$,
\begin{equation}
\frac{ < L >}{L_{class}} \sim_{a/L_{class} \rightarrow 0}
(a/L_{class})^{-\gamma}.
\end{equation}
By Eq.(2), $\gamma$ is related to $d_{H}$ via $d_{H}= 1 + \gamma$.
Because action (6) is parametrized in terms of the hopping parameter $\kappa$,
it is natural to consider another critical exponent $\alpha$ defined by
\begin{equation}
\frac{ < L > }{ L_{class} } \sim_{\kappa \rightarrow \kappa_{crit}}
\left( \frac{ \kappa_{crit} - \kappa } { \kappa_{crit} } \right)^{-\alpha}.
\end{equation}
The critical exponent $\gamma$ is defined in the continuum limit $a \rightarrow
0$. When $a$ goes to zero, the dimensionless lattice mass $ma$ goes to zero,
and $\kappa$ goes to its critical value $\kappa_{crit} = 1/8r$.
For a free Euclidean fermion theory, which we investigate numerically, both
exponents coincide,
$\alpha = \gamma$.

\begin{flushleft}
{\bf 4. Numerical results} \\
\end{flushleft}
Before measuring the length of the fermion propagator, it is useful to see if
the definition of the propagator length makes sense.
In order to have a meaningful length definition, one would expect $< L >$ to
obey a power law (11) when approaching $\kappa_{crit}$ for fixed $L_{class}$.
In order to get a first idea on the behavior of $< L >$ we have considered a
drastically simplified hopping matrix $M_{i,j} = \delta_{i,j+1}
+ \delta_{i+1,j}$. We have dropped any dependence from $\gamma$-matrices and
study the length as a function of $\kappa$. We have done numerical calculations
for $D=1$ with $\kappa_{crit}=1/2$ on lattices up to $N=120$
and for $D=2$ with $\kappa_{crit}=1/4$ on lattices up to $N=50$.
The numerical results confirm the expected scaling behavior
of Eq.(11).
the results yield $\alpha = 0.49$ for both $D=1$ and $D=2$.

\bigskip

Now we turn to the fermion propagator. For the free fermion case,
$\kappa_{crit} = 1/8r$ in $D=4$. The Euclidean free fermion propagator for
Wilson fermions is given in momentum space by \cite{kn:Mont94}
\begin{equation}
\tilde{ \Delta}_{k} = \left( 1 - 2 r \kappa \sum_{\mu=1}^{4} \cos k_{\mu} + 2
\kappa \sum_{\mu=1}^{4} i \gamma_{\mu} \sin k_{\mu} \right)^{-1}.
\end{equation}
This is related to the space-time propagator $\Delta_{x,y} \equiv < \psi_{x}
\bar{\psi}_{y} >$ by Fourier transformation
\begin{equation}
\Delta_{x,y} = \frac{1}{V} \sum_{k} e^{i k \cdot (x-y)} \tilde{ \Delta}_{k},
\end{equation}
where $V = N_{1} N_{2} N_{3} N_{4}$ is the lattice volume.
Note that $\tilde{\Delta}_{k}$ has a pole at $k=0$, $\kappa_{crit}=1/2Dr$ in
$D$ space-time dimensions.
We choose periodic or anti-periodic boundary conditions. They correspond to the
following choice of lattice momenta $k_{\mu}$ (see Ref.\cite{kn:Mont94}),
$ k_{\mu}= 2 \pi n_{\mu}/N_{\mu}$ corresponds to periodic boundary conditions
and $ k_{\mu}= 2 \pi (n_{\mu}+1/2)/N_{\mu}$ corresponds to anti-periodic
boundary conditions. In both cases, $n_{\mu} \in 0,1,\cdots,N_{\mu}-1$.
Thus in the anti-periodic case $k_{\mu}^{min} = \pi/N_{\mu}$ is the minimal
value of $k_{\mu}$.

\bigskip

We consider two components of the propagator: the unit-matrix component is
given by $\frac{1}{4} Tr[\Delta_{x,y}]$ and the $\gamma_{\mu}$-component by
$\frac{1}{4} Tr[\gamma_{\mu} \Delta_{x,y}]$.
Let us consider the cases of $D=1$, $D=2$ and $D=4$ space-time dimensions. For
$D=1$, one can compute the large lattice limit ($V \rightarrow \infty$, not the
continuum limit) analytically and obtains
\begin{eqnarray}
&& \frac{1}{4} Tr[\Delta_{x-y=n}] = 2^{n-1} \kappa^{n}, \;\;\; 1 \leq n, \;\;\;
\kappa \leq \kappa_{crit},
\nonumber \\
&& < L > = L_{class}.
\end{eqnarray}
However, we are interested in the continuum limit $a \rightarrow 0$, which
corresponds to $\kappa \rightarrow \kappa_{crit}$,
with $\kappa_{crit}=1/2r$ in $D=1$. The Wilson parameter is $r=1$. The
numerical results for the $\gamma_{1}$-component and the unit-component are
shown in Fig.[1a,b].
For the $\gamma_{1}$-component, we have chosen anti-periodic boundary
conditions. We have varied $N \equiv N_{1} = 4,8,16,\cdots,1024$.
In order to approach $\kappa_{crit}$ we have varied $\kappa = 1/[2 r \cos(k) +
2\sin(k)]$,
with $k = k^{min} = \pi/N$. We have chosen as classical length
$L_{class} = | x - y | = 1$. We have evaluated the space-time propagator
(13) by doing the Fourier transformation of $\tilde{\Delta}_{k}$ and of
$\frac{d}{d \kappa} \tilde{\Delta}_{k}$ numerically. From that we have
evaluated the length $< L >$ via Eq.(9) and hence the exponent $\alpha$ and
$d_{h}$ via Eq.(11). The result shows a power law behavior (11) with $\alpha =
-0.0016$, which corresponds to $d_{H}=0.9984$, by Eq.(2).

The behavior of the unit-component, with periodic boundary conditions, is
different. We have varied $N=20,40,\cdots,100$. Because for periodic boundary
conditions $k^{min}=0$,
we have approached $\kappa \rightarrow \kappa_{crit}$ by choosing $\kappa =
0.45, 0.475, 0.4875,\cdots$ (decreasing $| \kappa_{crit} - \kappa |$
by a factor 2 in each step.).
Now we have considered the classical length $L_{class} = N/2$. The computation
of $< L >$ and $d_{H}$ is as above.
One observes (Fig.[1b]) a power law with the critical exponent varying between
$\alpha =1.000$ and $\alpha = 0.9983$,
and the corresponding fractal dimension varying between $d_{H}=2.000$ and
$d_{H}=1.9983$.
Generally, one observes that the larger the size of the lattice, the closer one
has to be at $\kappa_{crit}$ before the scaling behavior (11) is seen.
How can a curve in topological dimension $D=1$ show a fractal dimension larger
than 1?
The Hausdorff dimension measures the hopping of the fermion forward and
backward on a line ($D=1$), which can be fractal.

\bigskip

Similar results are obtained in $D=2$ space-time dimensions, shown in
Fig.[2a,b]. Now $\kappa_{crit}=1/4r$.
Firstly, we have considered the $\gamma_{2}$-component.
Because we have a regular, symmetric lattice, the $k$-dependence is the same
for all $\gamma_{\mu}$-components.
Thus we can interpret the $\gamma_{1}$-component as space component and the
$\gamma_{2}$-component as time-component.
We have chosen boundary conditions periodic in space and anti-periodic in time.
We have varied $N_{1}=N_{space}=4,8,16$ independently from
$N_{2}=N_{time}=4,8,16,\cdots,1024$.
In order to approach $\kappa_{crit}$ we have varied
$\kappa = 1/[2 r(1 + \cos(k)) + 2\sin(k)]$,
where $k=k^{min}_{2} = \pi/N_{time}$.
As classical length we have chosen $L_{class}=N_{time}/2$.
We have obtained $\alpha=0.0022$ and $d_{H}=1.0022$.

For the unit-component, we have chosen periodic boundary conditions in space
and time. We have varied $N = N_{1} = N_{2} = 10,20,\cdots,50$. In order to
approach $\kappa_{crit}$
we have chosen $\kappa = 0.225, 0.2375, \cdots $
(decreasing $| \kappa_{crit} - \kappa |$ by a factor 2 in each step).
As classical length we have chosen $L_{class} = N/2$. We find $d_{H}=1.999$ to
$d_{h}=1.994$ for lattices varying between $N=10,\cdots,50$.

\bigskip

Finally, we present the results for $D=4$ in Fig.[3a,b]. Now
$\kappa_{crit}=1/8r$.
For the $\gamma_{4}$-component, we have chosen boundary conditions periodic in
space and anti-periodic in time.
We have varied $N_{1}=N_{2}=N_{3}=N_{space}=4,8,16$ and
$N_{4}=N_{time}=4,8,\cdots,256$.
In order to approach $\kappa_{crit}$ we have varied
$\kappa = 1/[2 r(3 + \cos(k)) + 2\sin(k)]$,
where $k=k^{min}_{4} = \pi/N_{time}$.
As classical length we have chosen $L_{class}=N_{time}/2$.
As results we obtain $\alpha=0.0086$ and $d_{H}=1.008$.

For the unit-component, we have chosen periodic boundary conditions in space
and time. We have varied $N = N_{1} = \cdots = N_{4} = 4,8,12\cdots,28$. In
order to approach $\kappa_{crit}$
we have chosen $\kappa = 0.100, 0.110, \cdots $
(decreasing $| \kappa_{crit} - \kappa |$ by a factor 2 in each step).
As classical length we have chosen $L_{class} = N/2$. We find $d_{H}=2.10$ to
$d_{h}=2.05$ for lattices varying between $N=4,\cdots,28$.

\bigskip

One obtains the following picture:
The $\gamma_{\mu}$-conmponent of the propagator shows results compatible with
$d_{H}=1$, i.e., no fractal behavior, in $D=1,2,4$ and different combinations
of periodic/anti-periodic boundary conditions. However, the unit-component
shows results compatible with $d_{H}=2$ for $D=1,2,4$, i.e., the same fractal
behavior as in non-relativistic quantum mechanics.
The numerical results, that is $d_{H}=1$ for the unit-component and $d_{H}=2$
for $\gamma_{\mu}$-component are independent of these boundary conditions.
For larger lattices, scaling sets in later (closer to $\kappa_{crit}$).
Numerical errors increase when approaching the singularity
$\kappa \rightarrow \kappa_{crit}$.
Also, numerical errors increase with the size of the lattice. Nevertheless, one
observes for $D=4$ that when increasing the lattice size, the Hausdorff
dimension moves closer to the value 2.

\bigskip

\begin{flushleft}
{\bf 5. Discussion} \\
\end{flushleft}
(a) Why differ the results of $d_{H}$ for different components?
Let us consider periodic boundary conditions and compare the unit-component
with the $\gamma_{1}$-component of the propagator. Then the propagator, given
in momentum space by Eq.(12), projects under the Fourier transformation (13)
onto the $\cos( k \cdot (x-y))$ part for the unit-component, but onto the $\sin
(k \cdot (x - y))$ part for the $\gamma_{1}$-component. In other words,
the unit- and the $\gamma_{1}$-component have a different pole structure when
$k \rightarrow 0$. Let us consider the continuum limit, $a \rightarrow 0$ and
$\epsilon = \kappa_{crit}- \kappa \rightarrow 0$, but keep the lattice volume
$V$ fixed. Also we keep $L_{class} = | x - y | = \mbox{const}.$ and consider in
$D=1$ the unit-component
$\frac{1}{4} Tr[\Delta_{x,y}]$, given by Eqs.(12,13), as a function of
$\epsilon$.
For small enough $\epsilon$, the dominant contributions come from lattice
momenta $k_{i}$ with $k_{i} << \epsilon$. Taking in Eq.(13) only those lattice
momenta into account, one finds
$\frac{1}{4} Tr[\Delta_{x,y}] \sim 1/\epsilon$.
Consequently,
$\kappa \frac{d}{d \kappa} \frac{1}{4} Tr[\Delta_{x,y}] \sim 1/\epsilon^{2}$.
Thus $< L > \sim 1/\epsilon$, which by Eq.(11) implies $d_{H}=2$.
Doing the analogous calculation for the $\gamma_{1}$-component yields
$\frac{1}{4} Tr[\gamma_{1} \Delta_{x,y}] \sim \mbox{const.}$
and
$\kappa \frac{d}{d \kappa} \frac{1}{4} Tr[\gamma_{1} \Delta_{x,y}] \sim
\mbox{const.}$. Thus $< L > \sim \mbox{const.}$ and hence $d_{H}=1$. This is in
agreement with our numerical results. \\
(b) The results for the free fermion propagator on the lattice can be compared
with Feynman's analytical expression for the fermion propagator in the
asymptotic regime $x^{2} << t^{2}$ and $x^{2} >> t^{2}$ \cite{kn:Feyn61}.
Feynman expresses the propagator kernel
$K_{+}(2,1) = i ( i \gamma_{\mu} \partial^{\mu} + m) I_{+}(t,\vec{x})$, and
gives for the function $I_{+}$ the asymptotic expression
\begin{equation}
I_{+}(t,\vec{x}) \rightarrow \exp \{ -i m[t -(x^{2}/2t)]\}, \;\;\;
x^{2} << t^{2},
\end{equation}
It can be seen that the propagation kernel is essentially the same as for a
free particle in non-relativistic quantum mechanics, where the Hausdorff
dimension is $d_{H}=2$.
This is in accord with our result $d_{H}=2$ for the unit-component of the
fermion propagator, which dominates the non-relativistic regime. \\
(c) The definition of length (9) for the action (6) can be generalized to the
case
when matter interacts with radiation, i.e., $QED$.
Then the fermion-photon interaction has the same structure as in Eq.(6), but
the hopping matrix $M[U]$ depends now on the gauge field via the link variables
$U_{\mu}(n)$
(for details see \cite{kn:Mont94}), and the matrix element $(K^{-1})_{m,n}$
occuring in
the fermion propagator, Eq.(8), must be replaced by a quantum expectation value
$< (K[U]^{-1})_{m,n} >$ which means doing a path integral over the gauge field.
However, one has to fix the gauge (see \cite{kn:Bern90}).

\bigskip

In summary, we have suggested a definition of length for the propagation of
relativistic fermions. It shows scaling behavior when approaching the continuum
limit and yields
the critical exponents $d_{H}=2$ (unit-component) and $d_{H}=1$
($\gamma_{\mu}$-component).
This is consistent with the analytical behavior of the Fermion propagator.
Our length definition can be directly generalized to interacting theories,
e.g., matter with radiation.

\begin{flushleft}
{\bf Acknowledgement}
\end{flushleft}
The author is grateful to J. Polonyi for very stimulating discussions on the
propagation length.
The author gratefully acknowledges support from NSERC Canada.

\newpage
\begin{flushleft}
{\bf Figure Caption}
\end{flushleft}
\begin{description}
\item[{Fig.1}]
$< L > / L_{class}$ versus $(\kappa_{crit} - \kappa)/\kappa_{crit}$ for free
fermion propagator in D=1,
(a) unit component, (b) $\gamma_{1}$-component of propagator.
\item[{Fig.2}]
Same as Fig.[1] in D=2 dimensions, (a) unit-component, (b)
$\gamma_{2}$-component.
\item[{Fig.3}]
Same as Fig.[1] in D=4 dimensions, (a) unit-component, (b)
$\gamma_{4}$-component.
\end{description}


\newpage
\begin{thebibliography}{999}
\bibitem{kn:Feyn65} R.P. Feynman and A.R. Hibbs,
Quantum Mechanics and Path Integrals, McGraw Hill, New York (1965).
\bibitem{kn:Mand83} B.B. Mandelbrot, The Fractal Geometry of Nature, Freeman,
New York (1983).
\bibitem{kn:Abbo81} L.F. Abbot and M.B. Wise,
Am. J. Phys. 49(1981)37.
\bibitem{kn:Nels66} E. Nelson, Phys. Rev. 150(1966)1079.
\bibitem{kn:Roep94} G. Roepstorff, Path Integral Approach to Quantum Mechanics,
Springer, New York (1994).
\bibitem{kn:Itzy89} C. Itzykson and J.B. Drouffe, Statistical Field Theory,
Cambridge Univ. Press, Cambridge (1989), Vol.I.
\bibitem{kn:Camp82} E. Campesino-Romeo, J.C.D'Olivio and M. Sokolovsky, Phys.
Lett. A89(1982)321.
\bibitem{kn:Krog95} H. Kr\"oger, S. Lantagne, K.J.M. Moriarty, and B. Plache,
Phys. Lett. A199(1995)299.
\bibitem{kn:Brue55} K.A. Brueckner, Phys. Rev. 97(1955)1353.
\bibitem{kn:Kawa90} H. Kawai and M. Ninomiya, Nucl. Phys. B336(1990)115.
\bibitem{kn:Davi92} F. David, Nucl. Phys. B368(1992)671.
\bibitem{kn:Kawa92} N. Kawamoto, V.A. Kazakov, Y. Saeki, Y. Watabiki, Phys.
Rev. Lett. 68(1992)2113.
\bibitem{kn:Kawa93} H. Kawai, N. Kawamoto, T. Mogami, Y. Watabiki, Phys. Lett.
B306(1993)19.
\bibitem{kn:Ambj94} J. Ambjorn , G. Thorleifson, Phys. Lett. B323(1994)7.
\bibitem{kn:Nish94} J. Nishimura, M. Oshikawa, Phys. Lett. B338(1994)187.
\bibitem{kn:Ambj95} J. Ambjorn , Y. Watabiki, Nils Bohr Inst. preprint
NBI-HE-95-01.
\bibitem{kn:Catt95} S. Catteral, G. Thorleifson, M. Bowick, V. John, Syracuse
Univ. preprint
(1995) SU-4240-607.
\bibitem{kn:Glim81} J. Glimm and A. Jaffe, Quantum Physics, Springer, New York
(1981).
\bibitem{kn:Oste72} K. Osterwalder, R. Schrader, Phys. Rev. Lett. 29(1972)1423;
Helv. Phys. Acta 46(1973)277; Comm. Math. Phys. 31(1973)83.
\bibitem{kn:Mont94} I. Montvay and G. M\"unster, Quantum Fields on a Lattice,
Cambridge Univ. Press, Cambridge (1994).
\bibitem{kn:Feyn61} R.P. Feynman, Quantum Electrodynamics, Benjamin, New York
(1961).
\bibitem{kn:Bern90} C. Bernard, D. Murphy, A. Soni, K. Yee, Nucl. Phys. B(Proc.
Suppl.)17(1990)593.
\end{thebibliography}
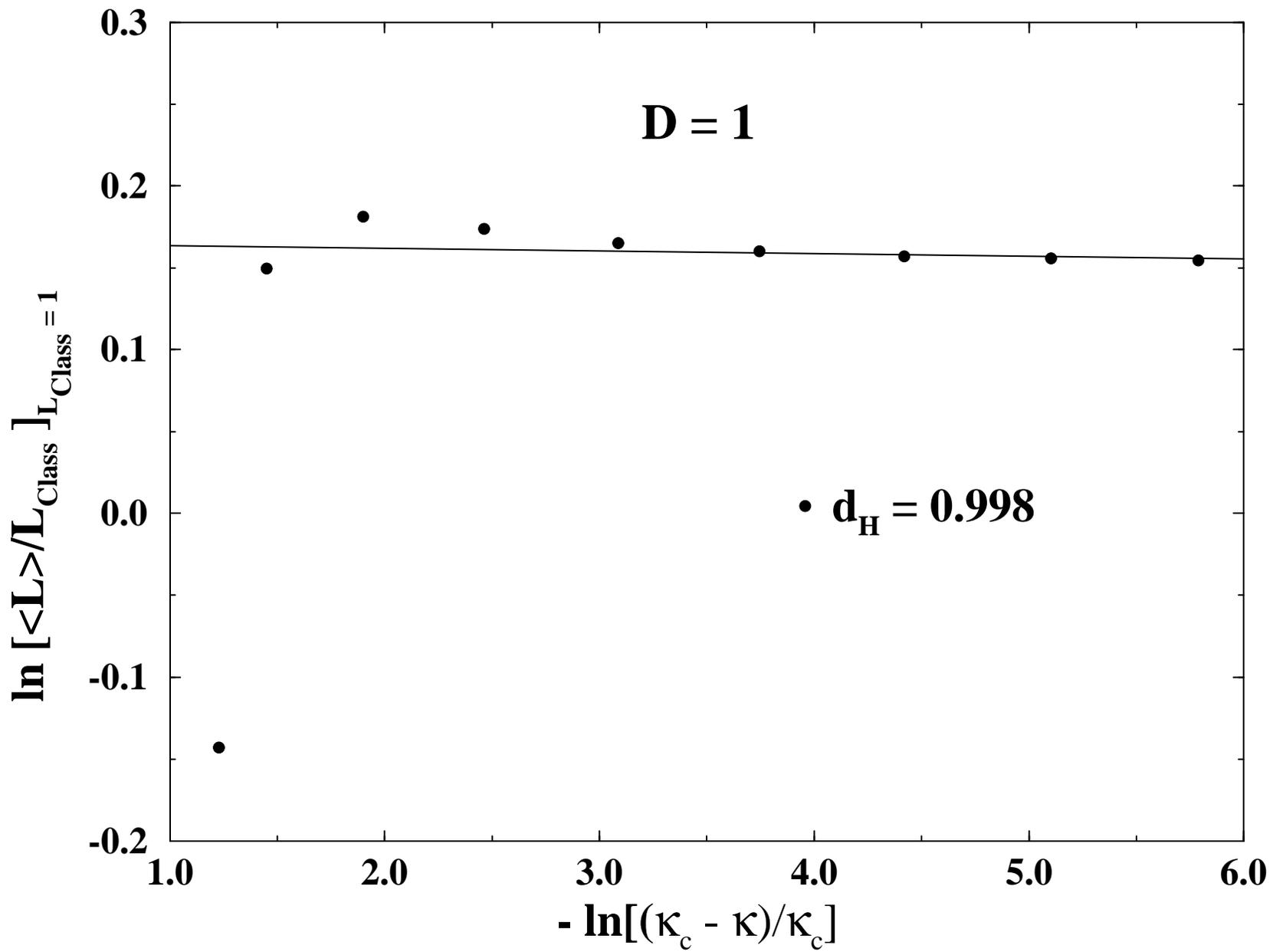
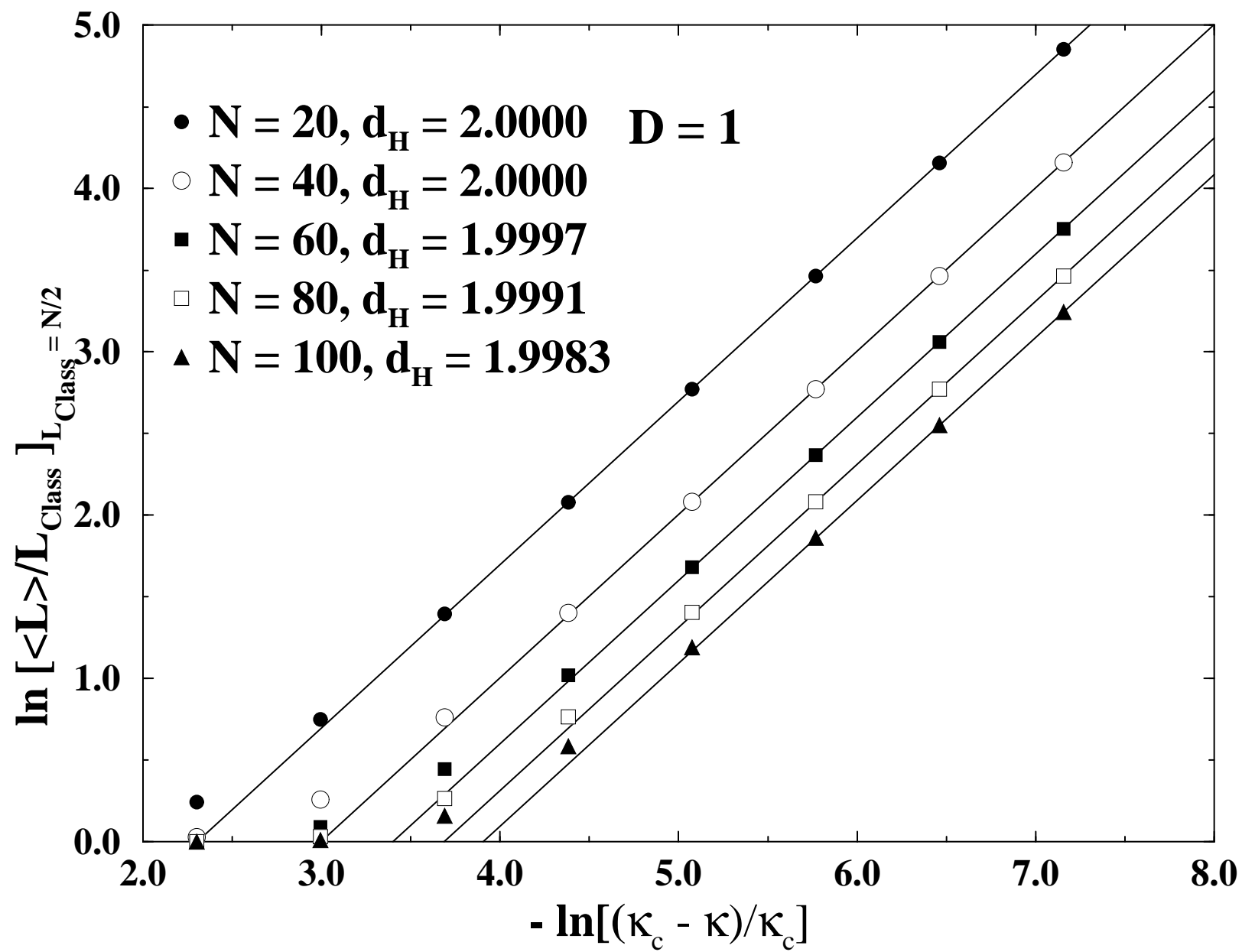
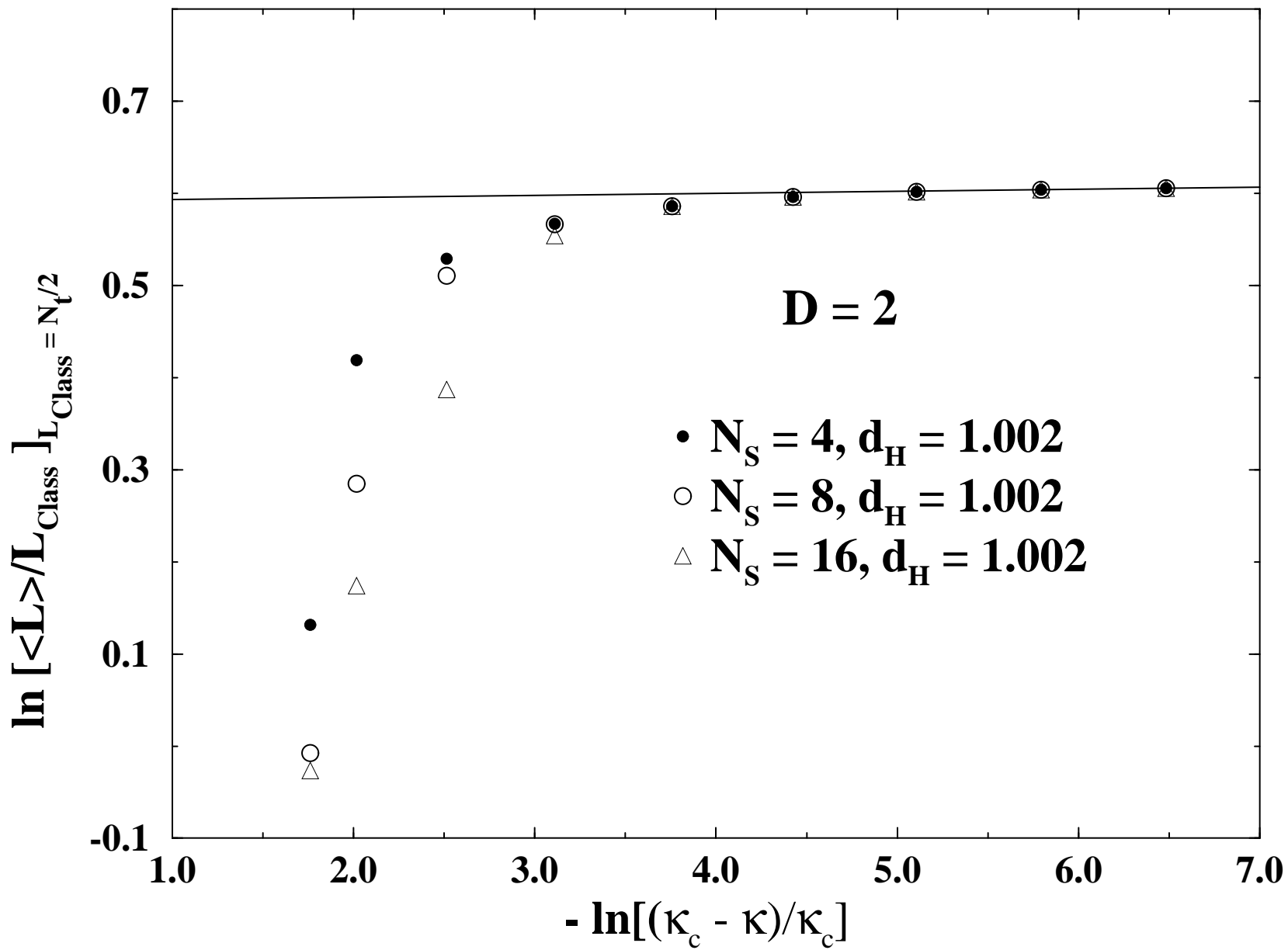
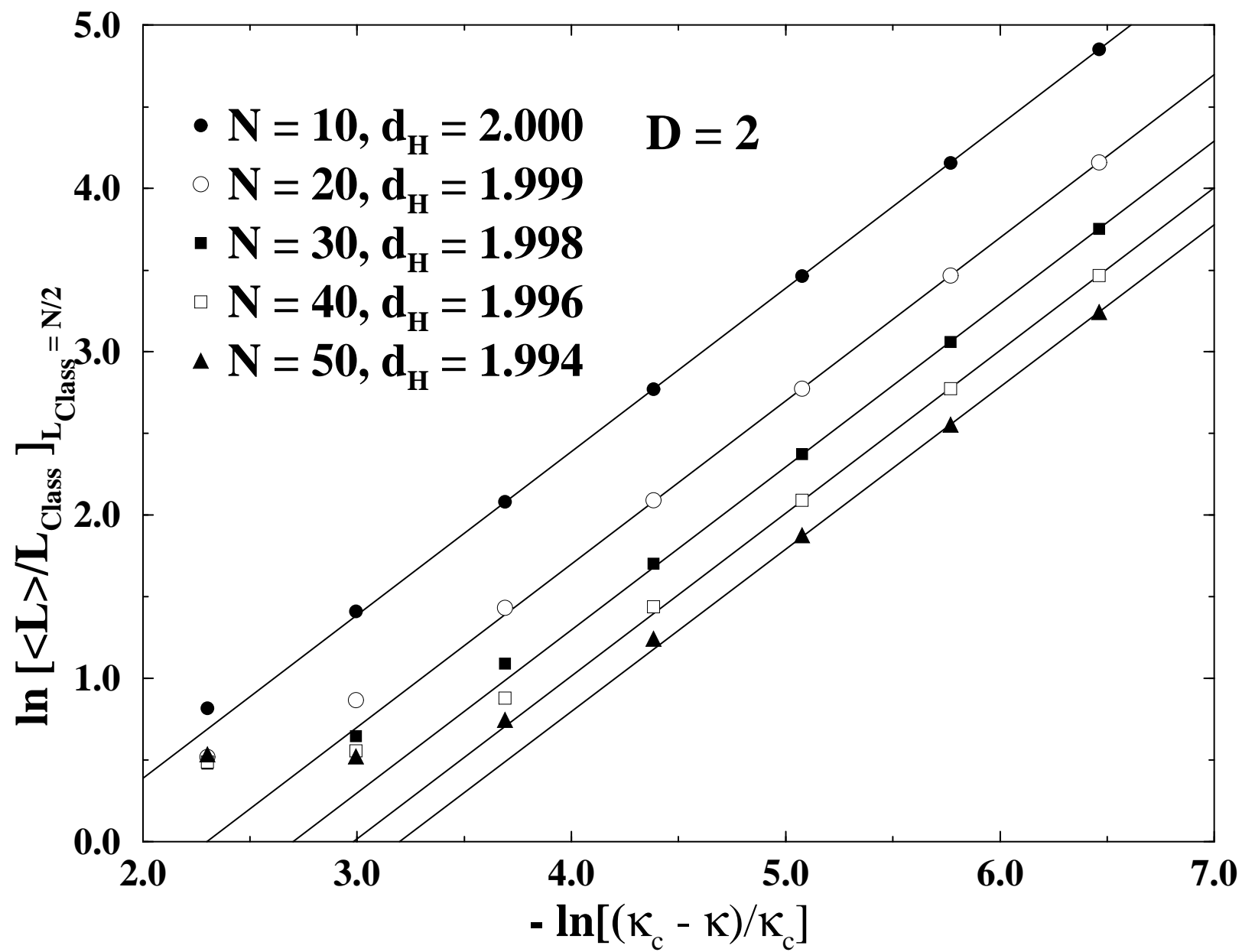
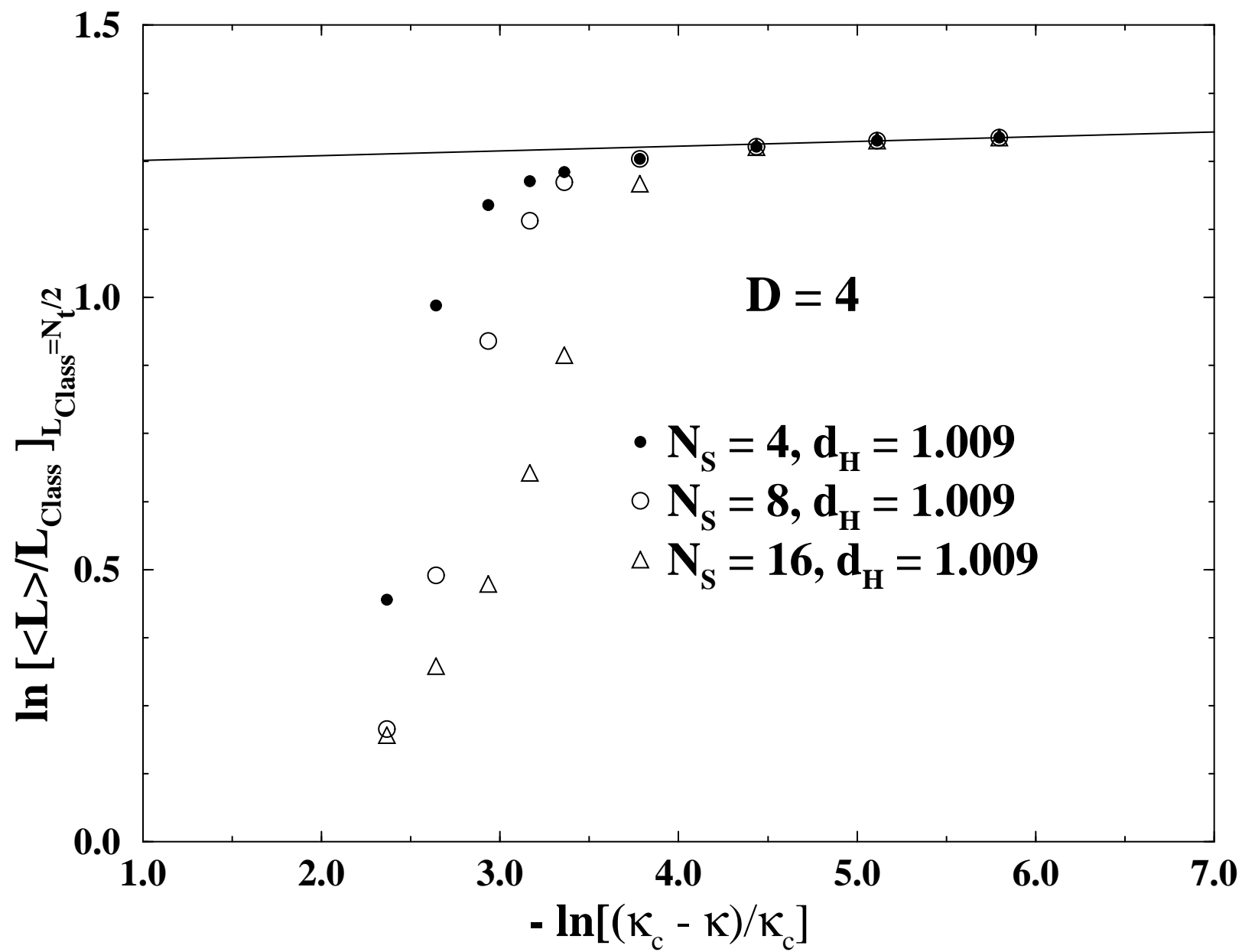
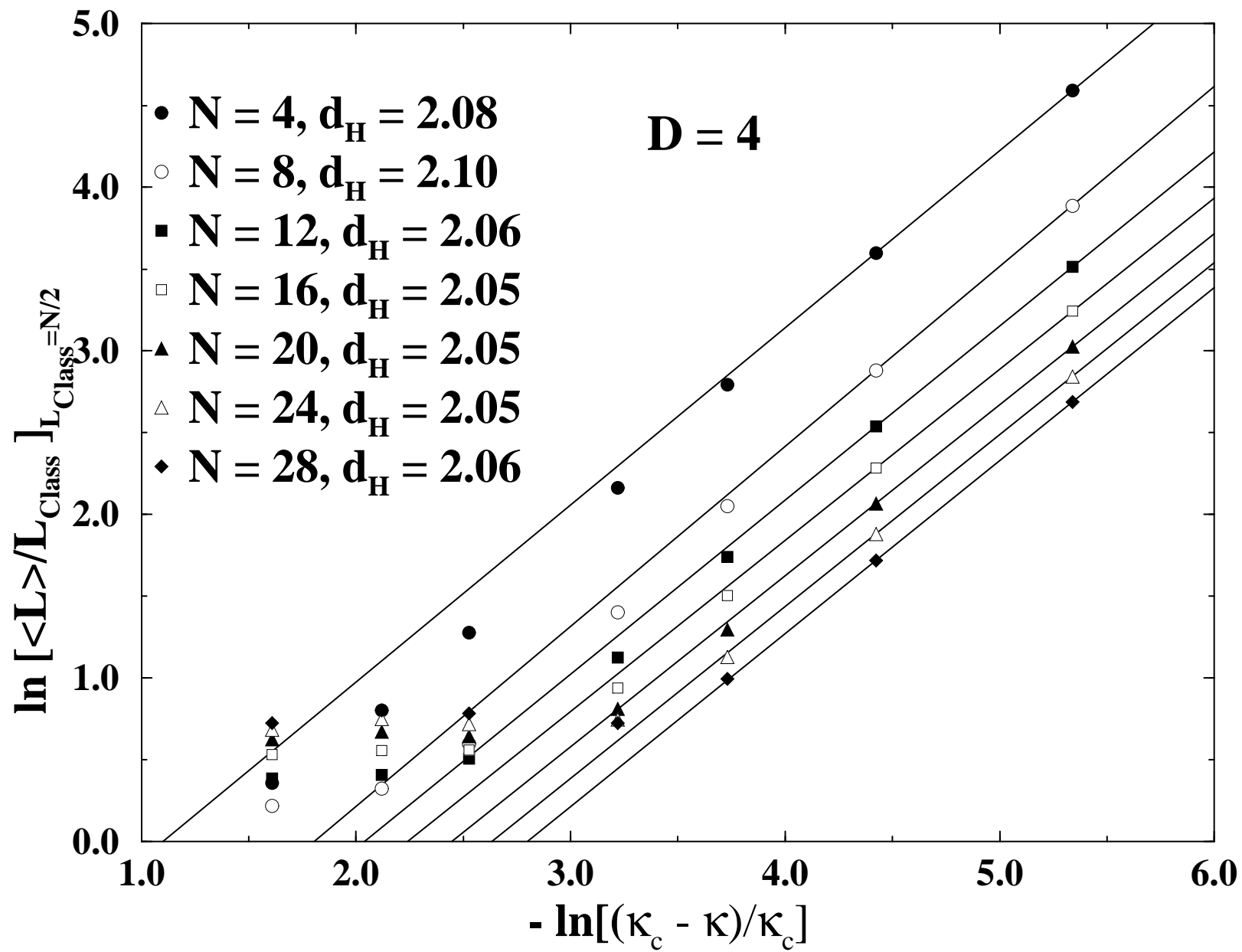
\end{document}